\documentclass[12pt]{article}

 \usepackage{graphicx}

\usepackage{amssymb}
\usepackage{amsmath}

\usepackage{txfonts}

\usepackage{latexsym}
\usepackage{theorem}

\usepackage{color,graphicx}

\newenvironment{proof}[1][Proof]
{\par\noindent{\bf #1:}}{\hspace*{\fill}\nolinebreak{$\Box$}\bigskip\par}

\newcommand{\qed}{\hspace*{\fill}\nolinebreak\ensuremath{\Box}}

\newtheorem{theorem}{Theorem}

\theorembodyfont{\upshape}

\begin{document}

\title{Phutball is PSPACE-hard}

\author{Dariusz Dereniowski\\
Department of Algorithms and System Modeling,\\
Gda\'{n}sk University of Technology, Poland\\
deren@eti.pg.gda.pl}

\maketitle

\begin{center}
\parbox[c]{10 cm}{
\textbf{Abstract:}
We consider the $n\times n$ game of Phutball. It is shown that,
given an arbitrary position of stones on the board, it is a
PSPACE-hard problem to determine whether the specified player
can win the game, regardless of the opponent's choices made
during the game.
}
\end{center}


\section{Introduction}

There is a deep mathematical theory developed for analyzing
combinatorial games \cite{Conway,numbers_and_games}.
The researchers work on the algorithmic techniques which are useful
for finding good game strategies for many board games, including
Phutball \cite{iterative_widening,admissible_moves}.
The paper \cite{generalized_threads} introduces the notion of generalized
threads and this technique is used to solve some Go instances and the
author suggest that this approach could be effective for other board
games, like Phutball. T.Cazenave used an approach called
Gradual Abstract Proof Search to show that $11\times 11$ Phutball is
a win for the first player \cite{gradual_abstract_proof}.
The game is \emph{loopy}, i.e. it is possible to obtain a configuration
of stones which already appeared in one of the previous turns -- some
combinatorial aspects of loopy games were considered in
\cite{loopy_games_phd}.
In this paper we are interested in the complexity of the game rather
than in manipulating and analyzing the rooted tree describing the game.
Several generalizations of one-player games turns out to be NP-complete:
Peg Solitaire \cite{solitaire-NPC},
Minesweeper (the problem of testing consistency) \cite{Minesweeper-NPC},
Same Game \cite{same_game-NPC}.
However, most of the board games (especially two-player games)
appear to be harder:
Checkers \cite{checkers_exptime},
Hex (a generalization to graphs) \cite{gen_of_hex},
Othello (Reversi) \cite{reversi_pspace},
Sokoban \cite{sokoban_pspace},
Go \cite{go_pspace,go_exptime,go_endgames},
Dyson Telescopes \cite{dyson_telescopes},
Rush Hour \cite{rush_hour_pspace} or
Amazons \cite{amazons_pspace}.

The Phutball \cite{Conway_phutball} game is usually played on a
$19\times 19$ Go board. Initially a black stone is placed in the middle
of the board. The players make their moves alternately. A player makes
his move by either placing a white stone in an unoccupied position, or
makes a sequence of \emph{jumps} over horizontal, vertical or
diagonal sequences of white stones. Each jump is performed by moving
the black stone, called \emph{ball}, over a line of white stones
(no empty space between the ball and the line is allowed if we want
to make a jump) and placing the ball on the board on the first unoccupied
position after the last white stone in the line. The white stones are
removed from the board immediately after the jump. Each player tries to
move the ball on or over the opponent's \emph{goal line}. The goal lines
are two opposite edges of the board.
We consider a natural generalization with an arbitrary size of the
board and initially a black stone placed in the middle of the board.

As indicated in \cite{Conway_phutball}, Phutball is not the kind of
game where you can expect a complete analysis. The authors in
\cite{one_dim_phutball} considered a simplified version of the
game, i.e. the case where there is only one dimension and it
turns out that according to the presented examples, the
one-dimensional version still seems to be hard to analyze. Moreover,
given an arbitrary position in the $2$-dimensional Phutball game, it is an NP-complete
problem to determine whether the current player can win the game
in his next move \cite{phutball_endgames}. However, as indicated in
several papers
\cite{gradual_abstract_proof,playing_games,phutball_endgames,Hearn_phd},
the complexity of the Phutball game is still open. In this paper we
place the problem of determining whether the current player has a
winning strategy in the class of PSPACE-hard problems.

\section{A graph game}
\label{sec:graph_game}

We start this section by describing the rules of a game played
on a graph. Then we prove that this game is PSPACE-hard.
The graph constructed on the basis of a problem
known to be PSPACE-complete is defined in such a way that
its topology allows to code it as a configuration
of stones in the Phutball game.

The game described in the following is played on a directed
graph. For completeness we list here some basic definitions.
A directed graph $G$ is a pair $G=(V(G),E(G))$ with a vertex set $V(G)$
and a set of directed edges $E(G)$ (each $e\in E(G)$ is an ordered
pair of two vertices). We say that $H$ is a \emph{subgraph} of $G$,
$H\subseteq G$, if $V(H)\subseteq V(G)$ and $E(H)\subseteq E(G)$.
A \emph{directed path} $P=(\{v_1,\ldots,v_n\},E(P))$ from $v_1$ to $v_n$
is a graph with edge set $E(P)=\{(v_i,v_{i+1}):i=1,\ldots,n-1\}$.
The vertices $V(P)\setminus\{v_1,v_n\}$ are the \emph{internal vertices}
of $P$.

The input of the game is a directed graph $G=(V(G),E(G))$,
a set $C\subseteq V(G)$, a vertex $s\in C$, and a relation
$R\subseteq V(G)\times E(G)$ between the vertices and the edges of $G$.
If $(v,e)\in R$ then we say that a vertex $v$
is \emph{pointing} an edge $e$. Denote by $R^{-1}(E(G))$ the set of
vertices $v$ for which there exists $e\in E(G)$, such that $(v,e)\in R$.
The players of the game will be called $\exists$-\emph{player}
and $\forall$-\emph{player}. We will also use a notation that if
a symbol $X$ refers to one of the players then $\overline{X}$ is the
other player.

At each point of the game there is a unique \emph{active vertex}.
The players must follow the rules:
\begin{list}{Rule}{}
\item{$1$} (\emph{initialization}). The $\exists$-player starts the game.
     Initially $s$ is the active vertex.
\item{$2$} (\emph{a move}). Let $u\in C$ be the active
     vertex. The current player $X$ selects a vertex
     $v\in C\cup R^{-1}(E(G))$
     and a directed path $P\subseteq G$ from $u$ to $v$
     such that all internal vertices of $P$ are in
     $V(G)\setminus (C\cup R^{-1}(E(G)))$.
     The edges of $P$ are removed from $G$, $v$
     becomes the active vertex, and $\overline{X}$ becomes the
     current player. We say that $X$ \emph{moves from $u$ to $v$}.
\item{$3$} (\emph{game end}). If the current player cannot
     make a move, i.e. there is no directed path $P$ from the
     active vertex to a vertex $v\in C\cup R^{-1}(E(G))$,
     then the current player loses the game.
     If the current player moves from $u$ to $v\in R^{-1}(E(G))$
     then he wins the game.
\end{list}

Let us recall the PSPACE-complete Quantified Boolean Formula
(\emph{QBF}) problem \cite{QBF}.
Given a formula $Q$ in the form
\[Q_1x_1 \cdots Q_nx_n F(x_1,\ldots,x_n),\]
decide whether the formula is true, where $Q_i\in\{\exists,\forall\}$
for $i=1,\ldots,n$.
In our case we us a restricted case of this problem where
$Q_1=\exists$, $Q_{i+1}\neq Q_i$ for $i=1,\ldots,n-1$,
the integer $n$ is even, and $F$ is a 3CNF formula, i.e.
$F=F_1\land F_2\land\cdots\land F_m$, where
$F_i=(l_{i,1}\lor l_{i,2}\lor l_{i,3})$ and each literal
$l_{i,j}$ is a variable or the negation of a variable,
$i=1,\ldots,m,j=1,2,3$.

Given $Q$, we create a directed graph $G$. For each variable
$x_i$ define the corresponding variable component $G(x_i)$:
\[V(G(x_i))=\{a_i,b_i,c_i,d_i,e_i,f_i,g_i\},\]
\[E(G(x_i))=\{(a_i,b_i),(a_i,c_i),(b_i,e_i),(c_i,f_i),(e_i,d_i),(f_i,d_i),(d_i,g_i)\},\]
for $i=1,\ldots,n$.
We connect the variable components in such a way that
$(g_i,a_{i+1})\in E(G)$ for each $i=1,\ldots,n-1$.

\begin{figure}[htb]
\begin{center}
\begin{picture}(0,0)%
\includegraphics{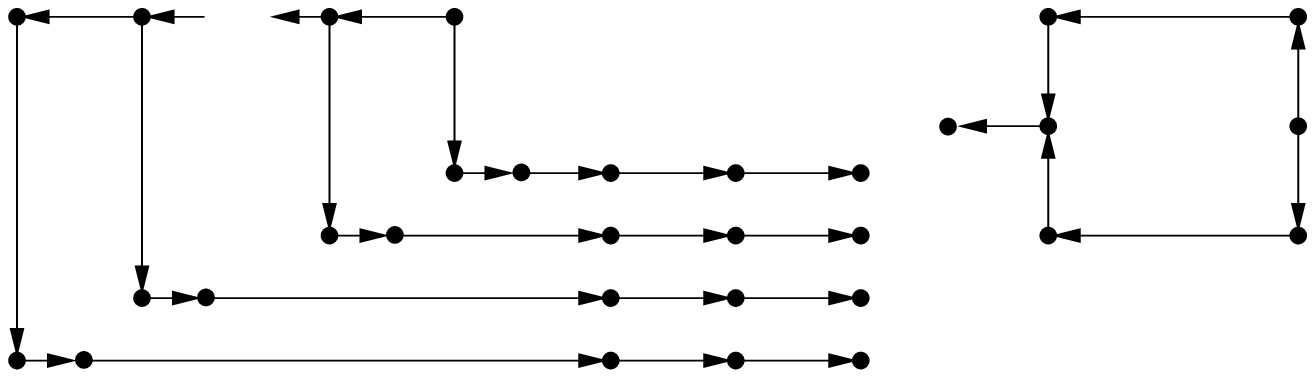}%
\end{picture}%
\setlength{\unitlength}{3947sp}%
\begingroup\makeatletter\ifx\SetFigFont\undefined%
\gdef\SetFigFont#1#2#3#4#5{%
  \reset@font\fontsize{#1}{#2pt}%
  \fontfamily{#3}\fontseries{#4}\fontshape{#5}%
  \selectfont}%
\fi\endgroup%
\begin{picture}(6480,2374)(61,-1625)
\put(6526,-286){\makebox(0,0)[lb]{\smash{{\SetFigFont{10}{12.0}{\familydefault}{\mddefault}{\updefault}{\color[rgb]{0,0,0}$a_i$}%
}}}}
\put(6526,314){\makebox(0,0)[lb]{\smash{{\SetFigFont{10}{12.0}{\familydefault}{\mddefault}{\updefault}{\color[rgb]{0,0,0}$b_i$}%
}}}}
\put(1201,164){\makebox(0,0)[lb]{\smash{{\SetFigFont{10}{12.0}{\familydefault}{\mddefault}{\updefault}{\color[rgb]{0,0,0}$\ldots$}%
}}}}
\put(1201,-961){\makebox(0,0)[lb]{\smash{{\SetFigFont{10}{12.0}{\familydefault}{\mddefault}{\updefault}{\color[rgb]{0,0,0}$\ldots$}%
}}}}
\put(2476,314){\makebox(0,0)[lb]{\smash{{\SetFigFont{10}{12.0}{\familydefault}{\mddefault}{\updefault}{\color[rgb]{0,0,0}$x_1$}%
}}}}
\put(2176,-511){\makebox(0,0)[lb]{\smash{{\SetFigFont{10}{12.0}{\familydefault}{\mddefault}{\updefault}{\color[rgb]{0,0,0}$y_1$}%
}}}}
\put( 76,-1486){\makebox(0,0)[lb]{\smash{{\SetFigFont{10}{12.0}{\familydefault}{\mddefault}{\updefault}{\color[rgb]{0,0,0}$y_m$}%
}}}}
\put(2551,-661){\makebox(0,0)[lb]{\smash{{\SetFigFont{10}{12.0}{\familydefault}{\mddefault}{\updefault}{\color[rgb]{0,0,0}$z_1$}%
}}}}
\put(2026,-961){\makebox(0,0)[lb]{\smash{{\SetFigFont{10}{12.0}{\familydefault}{\mddefault}{\updefault}{\color[rgb]{0,0,0}$z_2$}%
}}}}
\put(1126,-1261){\makebox(0,0)[lb]{\smash{{\SetFigFont{10}{12.0}{\familydefault}{\mddefault}{\updefault}{\color[rgb]{0,0,0}$z_{m-1}$}%
}}}}
\put(526,-1561){\makebox(0,0)[lb]{\smash{{\SetFigFont{10}{12.0}{\familydefault}{\mddefault}{\updefault}{\color[rgb]{0,0,0}$z_m$}%
}}}}
\put(526,-1186){\makebox(0,0)[lb]{\smash{{\SetFigFont{10}{12.0}{\familydefault}{\mddefault}{\updefault}{\color[rgb]{0,0,0}$y_{m-1}$}%
}}}}
\put(1576,-886){\makebox(0,0)[lb]{\smash{{\SetFigFont{10}{12.0}{\familydefault}{\mddefault}{\updefault}{\color[rgb]{0,0,0}$y_2$}%
}}}}
\put(1726,314){\makebox(0,0)[lb]{\smash{{\SetFigFont{10}{12.0}{\familydefault}{\mddefault}{\updefault}{\color[rgb]{0,0,0}$x_2$}%
}}}}
\put(826,314){\makebox(0,0)[lb]{\smash{{\SetFigFont{10}{12.0}{\familydefault}{\mddefault}{\updefault}{\color[rgb]{0,0,0}$x_{m-1}$}%
}}}}
\put(151,314){\makebox(0,0)[lb]{\smash{{\SetFigFont{10}{12.0}{\familydefault}{\mddefault}{\updefault}{\color[rgb]{0,0,0}$x_m$}%
}}}}
\put(3001,-436){\makebox(0,0)[lb]{\smash{{\SetFigFont{10}{12.0}{\familydefault}{\mddefault}{\updefault}{\color[rgb]{0,0,0}$w_{1,1}$}%
}}}}
\put(3001,-736){\makebox(0,0)[lb]{\smash{{\SetFigFont{10}{12.0}{\familydefault}{\mddefault}{\updefault}{\color[rgb]{0,0,0}$w_{2,1}$}%
}}}}
\put(3526,-736){\makebox(0,0)[lb]{\smash{{\SetFigFont{10}{12.0}{\familydefault}{\mddefault}{\updefault}{\color[rgb]{0,0,0}$w_{2,2}$}%
}}}}
\put(4201,-736){\makebox(0,0)[lb]{\smash{{\SetFigFont{10}{12.0}{\familydefault}{\mddefault}{\updefault}{\color[rgb]{0,0,0}$w_{2,3}$}%
}}}}
\put(3001,-1036){\makebox(0,0)[lb]{\smash{{\SetFigFont{10}{12.0}{\familydefault}{\mddefault}{\updefault}{\color[rgb]{0,0,0}$w_{m-1,1}$}%
}}}}
\put(3601,-1036){\makebox(0,0)[lb]{\smash{{\SetFigFont{10}{12.0}{\familydefault}{\mddefault}{\updefault}{\color[rgb]{0,0,0}$w_{m-1,2}$}%
}}}}
\put(4201,-1036){\makebox(0,0)[lb]{\smash{{\SetFigFont{10}{12.0}{\familydefault}{\mddefault}{\updefault}{\color[rgb]{0,0,0}$w_{m-1,3}$}%
}}}}
\put(3001,-1336){\makebox(0,0)[lb]{\smash{{\SetFigFont{10}{12.0}{\familydefault}{\mddefault}{\updefault}{\color[rgb]{0,0,0}$w_{m,1}$}%
}}}}
\put(3601,-1336){\makebox(0,0)[lb]{\smash{{\SetFigFont{10}{12.0}{\familydefault}{\mddefault}{\updefault}{\color[rgb]{0,0,0}$w_{m,2}$}%
}}}}
\put(4201,-1336){\makebox(0,0)[lb]{\smash{{\SetFigFont{10}{12.0}{\familydefault}{\mddefault}{\updefault}{\color[rgb]{0,0,0}$w_{m,3}$}%
}}}}
\put(6526,-961){\makebox(0,0)[lb]{\smash{{\SetFigFont{10}{12.0}{\familydefault}{\mddefault}{\updefault}{\color[rgb]{0,0,0}$c_i$}%
}}}}
\put(5026,-961){\makebox(0,0)[lb]{\smash{{\SetFigFont{10}{12.0}{\familydefault}{\mddefault}{\updefault}{\color[rgb]{0,0,0}$f_i$}%
}}}}
\put(5326,-286){\makebox(0,0)[lb]{\smash{{\SetFigFont{10}{12.0}{\familydefault}{\mddefault}{\updefault}{\color[rgb]{0,0,0}$d_i$}%
}}}}
\put(5026,314){\makebox(0,0)[lb]{\smash{{\SetFigFont{10}{12.0}{\familydefault}{\mddefault}{\updefault}{\color[rgb]{0,0,0}$e_i$}%
}}}}
\put(4576,-211){\makebox(0,0)[lb]{\smash{{\SetFigFont{10}{12.0}{\familydefault}{\mddefault}{\updefault}{\color[rgb]{0,0,0}$g_i$}%
}}}}
\put(151,614){\makebox(0,0)[lb]{\smash{{\SetFigFont{10}{12.0}{\familydefault}{\mddefault}{\updefault}{\color[rgb]{0,0,0}($a$)}%
}}}}
\put(4576,614){\makebox(0,0)[lb]{\smash{{\SetFigFont{10}{12.0}{\familydefault}{\mddefault}{\updefault}{\color[rgb]{0,0,0}($b$)}%
}}}}
\put(3601,-436){\makebox(0,0)[lb]{\smash{{\SetFigFont{10}{12.0}{\familydefault}{\mddefault}{\updefault}{\color[rgb]{0,0,0}$w_{1,2}$}%
}}}}
\put(4201,-436){\makebox(0,0)[lb]{\smash{{\SetFigFont{10}{12.0}{\familydefault}{\mddefault}{\updefault}{\color[rgb]{0,0,0}$w_{1,3}$}%
}}}}
\end{picture}%
\caption{The graphs ($a$) $G(F)$ and ($b$) $G(x_i)$}
\label{pic:components}
\end{center}
\end{figure}

Then we define the formula component $G(F)$:
\[V(G(F))=\{x_i,y_i,z_i:i=1,\ldots,m\}\cup\{w_{i,j}:i=1,\ldots,m,j=1,2,3\},\]
\begin{multline}
E(G(F))=\{(x_i,y_i),(y_i,z_i),(z_i,w_{i,1}),(w_{i,1},w_{i,2}),(w_{i,2},w_{i,3}):i=1,\ldots,m\}\cup\\
\{(x_i,x_{i+1}):i=1,\ldots,m-1\}.\nonumber
\end{multline}
Fig.~\ref{pic:components}($a$)
shows the formula component while Fig.~\ref{pic:components}($b$)
gives the variable component.
To finish the construction of $G$ let $(g_n,x_1)\in E(G)$ and
introduce a vertex $g_0$ connected
to the graph in such a way that $(g_0,a_1)\in E(G)$.

The input to our graph game is the directed graph $G$ defined above,
the set $C=\{g_0,\ldots,g_{n-1}\}\cup\{z_1,\ldots,z_m\}$,
$s=g_0$ and $R$ containing a pair $(w_{i,j},(b_l,e_l))$
(respectively $(w_{i,j},(c_l,f_l))$) for $i\in\{1,\ldots,m\}$, $j\in\{1,2,3\}$,
$l\in\{1,\ldots,n\}$, iff $l_{i,j}=x_l$ ($l_{i,j}=\overline{x_l}$, resp.).
Observe that initially $R^{-1}(E(G))$ contains all the vertices
$w_{i,j}$, $i=1,\ldots,m$, $j=1,2,3$, because a vertex $w_{i,j}$
corresponds to the literal $l_{i,j}$, which equals $x_l$ or $\overline{x_l}$
for some $l\in\{1,\ldots,n\}$. However, during the game the set $R^{-1}$
gets smaller due to the fact that some of the edges of $G$ are removed from $G$.

\begin{figure}[htb]
\begin{center}
\begin{picture}(0,0)%
\includegraphics{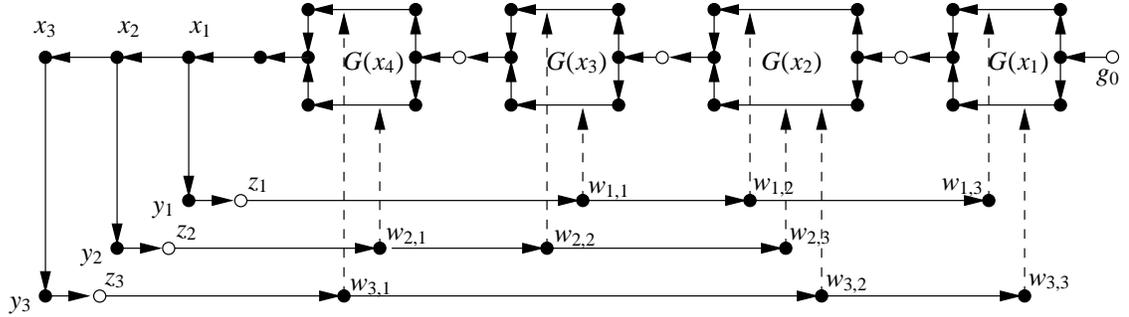}%
\end{picture}%
\setlength{\unitlength}{3947sp}%
\begingroup\makeatletter\ifx\SetFigFont\undefined%
\gdef\SetFigFont#1#2#3#4#5{%
  \reset@font\fontsize{#1}{#2pt}%
  \fontfamily{#3}\fontseries{#4}\fontshape{#5}%
  \selectfont}%
\fi\endgroup%
\begin{picture}(6988,1984)(361,-1700)
\put(1501, 89){\makebox(0,0)[lb]{\smash{{\SetFigFont{10}{12.0}{\familydefault}{\mddefault}{\updefault}{\color[rgb]{0,0,0}$x_1$}%
}}}}
\put(826,-1336){\makebox(0,0)[lb]{\smash{{\SetFigFont{10}{12.0}{\familydefault}{\mddefault}{\updefault}{\color[rgb]{0,0,0}$y_2$}%
}}}}
\put(1051, 89){\makebox(0,0)[lb]{\smash{{\SetFigFont{10}{12.0}{\familydefault}{\mddefault}{\updefault}{\color[rgb]{0,0,0}$x_2$}%
}}}}
\put(526, 89){\makebox(0,0)[lb]{\smash{{\SetFigFont{10}{12.0}{\familydefault}{\mddefault}{\updefault}{\color[rgb]{0,0,0}$x_3$}%
}}}}
\put(1276,-1036){\makebox(0,0)[lb]{\smash{{\SetFigFont{10}{12.0}{\familydefault}{\mddefault}{\updefault}{\color[rgb]{0,0,0}$y_1$}%
}}}}
\put(5101,-136){\makebox(0,0)[lb]{\smash{{\SetFigFont{10}{12.0}{\familydefault}{\mddefault}{\updefault}{\color[rgb]{0,0,0}$G(x_2)$}%
}}}}
\put(2476,-136){\makebox(0,0)[lb]{\smash{{\SetFigFont{10}{12.0}{\familydefault}{\mddefault}{\updefault}{\color[rgb]{0,0,0}$G(x_4)$}%
}}}}
\put(3751,-136){\makebox(0,0)[lb]{\smash{{\SetFigFont{10}{12.0}{\familydefault}{\mddefault}{\updefault}{\color[rgb]{0,0,0}$G(x_3)$}%
}}}}
\put(6526,-136){\makebox(0,0)[lb]{\smash{{\SetFigFont{10}{12.0}{\familydefault}{\mddefault}{\updefault}{\color[rgb]{0,0,0}$G(x_1)$}%
}}}}
\put(376,-1636){\makebox(0,0)[lb]{\smash{{\SetFigFont{10}{12.0}{\familydefault}{\mddefault}{\updefault}{\color[rgb]{0,0,0}$y_3$}%
}}}}
\put(1426,-1186){\makebox(0,0)[lb]{\smash{{\SetFigFont{10}{12.0}{\familydefault}{\mddefault}{\updefault}{\color[rgb]{0,0,0}$z_2$}%
}}}}
\put(7201,-211){\makebox(0,0)[lb]{\smash{{\SetFigFont{10}{12.0}{\familydefault}{\mddefault}{\updefault}{\color[rgb]{0,0,0}$g_0$}%
}}}}
\put(1876,-886){\makebox(0,0)[lb]{\smash{{\SetFigFont{10}{12.0}{\familydefault}{\mddefault}{\updefault}{\color[rgb]{0,0,0}$z_1$}%
}}}}
\put(976,-1486){\makebox(0,0)[lb]{\smash{{\SetFigFont{10}{12.0}{\familydefault}{\mddefault}{\updefault}{\color[rgb]{0,0,0}$z_3$}%
}}}}
\put(2516,-1526){\makebox(0,0)[lb]{\smash{{\SetFigFont{10}{12.0}{\familydefault}{\mddefault}{\updefault}{\color[rgb]{0,0,0}$w_{3,1}$}%
}}}}
\put(3798,-1221){\makebox(0,0)[lb]{\smash{{\SetFigFont{10}{12.0}{\familydefault}{\mddefault}{\updefault}{\color[rgb]{0,0,0}$w_{2,2}$}%
}}}}
\put(4010,-915){\makebox(0,0)[lb]{\smash{{\SetFigFont{10}{12.0}{\familydefault}{\mddefault}{\updefault}{\color[rgb]{0,0,0}$w_{1,1}$}%
}}}}
\put(5046,-894){\makebox(0,0)[lb]{\smash{{\SetFigFont{10}{12.0}{\familydefault}{\mddefault}{\updefault}{\color[rgb]{0,0,0}$w_{1,2}$}%
}}}}
\put(5275,-1208){\makebox(0,0)[lb]{\smash{{\SetFigFont{10}{12.0}{\familydefault}{\mddefault}{\updefault}{\color[rgb]{0,0,0}$w_{2,3}$}%
}}}}
\put(5503,-1501){\makebox(0,0)[lb]{\smash{{\SetFigFont{10}{12.0}{\familydefault}{\mddefault}{\updefault}{\color[rgb]{0,0,0}$w_{3,2}$}%
}}}}
\put(6232,-895){\makebox(0,0)[lb]{\smash{{\SetFigFont{10}{12.0}{\familydefault}{\mddefault}{\updefault}{\color[rgb]{0,0,0}$w_{1,3}$}%
}}}}
\put(6778,-1493){\makebox(0,0)[lb]{\smash{{\SetFigFont{10}{12.0}{\familydefault}{\mddefault}{\updefault}{\color[rgb]{0,0,0}$w_{3,3}$}%
}}}}
\put(2741,-1208){\makebox(0,0)[lb]{\smash{{\SetFigFont{10}{12.0}{\familydefault}{\mddefault}{\updefault}{\color[rgb]{0,0,0}$w_{2,1}$}%
}}}}
\end{picture}%
\caption{A complete instance of the graph $G$ corresponding to the
         formula in (\ref{eq:QBF_example})}
\label{pic:graph_ex}
\end{center}
\end{figure}

All the edges of the graph have vertical and horizontal orientations
(as shown in Figs~\ref{pic:components} and~\ref{pic:graph_ex})
and the lines of stones in the Phutball game corresponding to the
edges of the graph will preserve this topology.

Let us consider the following complete example of our reduction.
Given a formula $Q$
\begin{equation}
\exists x_1 \forall x_2 \exists x_3 \forall x_4
(x_1\lor x_2\lor\overline{x_3})\land
(\overline{x_2}\lor x_3\lor\overline{x_4})\land
(\overline{x_1}\lor\overline{x_2}\lor x_4),
\label{eq:QBF_example}
\end{equation}
Fig.~\ref{pic:graph_ex} depicts the corresponding graph $G$. The dashed arcs
represent the elements of the relation $R$, the vertices in $C$
are denoted as white nodes, while the vertices in $V(G)\setminus C$ are
the black nodes. Note that for each vertex $w_{i,j}$ there is exactly
one element $(w_{i,j},e)\in R$.
Since the $\exists$-player starts the game and the active vertex is $g_0$,
two moves are possible, i.e. $V(P)$ contains $g_0,a_1,b_1,e_1,d_1,g_1$ or
$g_0,a_1,c_1,f_1,d_1,g_1$. After the move $w_{1,3}$ or $w_{3,3}$ does
not belong to the set $R^{-1}(E(G))$, respectively.

The game obtained in the reduction has a special structure,
which makes it quite easy to analyze.
Here we list three straightforward facts describing the structure
of the game.
\begin{list}{\textbf{Fact}}{}
\item{\textbf{1}}
  If the active vertex is $g_{i-1}$,
  $i\in\{1,\ldots,n\}$ then the current player makes a move
  from $g_{i-1}$ to $g_i$ and the directed path $P$ removed
  from $G$ contains one of the following
  sequences of edges:
  \begin{equation}
    (g_{i-1},a_i),(a_i,b_i),(b_i,e_i),(e_i,d_i),(d_i,g_i),
    \label{eq:seq1}
  \end{equation}
  \begin{equation}
    (g_{i-1},a_i),(a_i,c_i),(c_i,f_i),(f_i,d_i),(d_i,g_i).
    \label{eq:seq2}
  \end{equation}
  Furthermore, the $\exists$-player makes such a move for $i=1,3,5\ldots,n-1$
  while the $\forall$-player makes this move for $i=2,4,6,\ldots,n$
  (recall that $n$ is even).
  $\qed$
\item{\textbf{2}}
  Let $g_n$ be the active vertex.
  The $\forall$-player is the current player and he makes a move
  from $g_n$ to a vertex $z_i$
  and the path $P$ contains the edges
  \begin{equation}
    (g_n,x_1),(x_1,x_2),\ldots,(x_{i-1},x_i),(x_i,y_i),(y_i,z_i),
    \label{eq:seq3}
  \end{equation}
  where $i\in\{1,\ldots,m\}$.
  $\qed$
\item{\textbf{3}}
  Assume that $z_i$, $i\in\{1,\ldots,m\}$, is the current vertex.
  The $\exists$-player is the current player.
  If there exists such a vertex $w_{i,j}$ that
  $(w_{i,j},e)\in R$ and $e\in E(G)$ then
  $\exists$-player wins the game. Otherwise he cannot make a move
  and he loses the game.
  $\qed$
\end{list}

\begin{theorem}
The above graph game is \textup{PSPACE}-hard.
\label{thm:graph_game_pspace}
\end{theorem}
\begin{proof}
We show that the $\exists$-player has a winning strategy if and
only if the corresponding quantified formula $Q$ is true. If we write
$e\in E(G)$ then we mean that $e$ is an edge of $G$ at the current
stage of the game.

First, assume
that the formula is true. Define the strategy as follows. If the
active vertex is $g_{i-1}$ and
the variable $x_i$, $i\in\{1,3,5,\ldots,n-1\}$ is true (in an assignment
of Boolean values to the variables forcing $F$ to be true) then
the $\exists$-player traverses the sequence of edges as stated
in (\ref{eq:seq2}). Otherwise the $\exists$-player traverses the
edges listed in (\ref{eq:seq1}). By Fact~2, the $\forall$-player
chooses in his last move an index $i\in\{1,\ldots,m\}$ and ends the move at
a vertex $z_i$. Since $F$ is true, $F_i$ is true, and consequently,
there exists a literal
$l_{i,j}$, $j\in\{1,2,3\}$, which is true. If $l_{i,j}=x_l$
for some $l\in\{1,\ldots,n\}$, then $x_l$ is true which implies that
$(b_l,e_l)\in E(G)$. Moreover, by the definition of $R$,
$w_{i,j}$ is pointing $(b_l,e_l)$. Similarly, if $l_{i,j}=\overline{x_l}$
for $l\in\{1,\ldots,n\}$, then $x_l$ is false, so $(c_i,f_i)\in E(G)$
and $(w_{i,j},(c_i,f_i))\in R$. So, for each choice of $i$ by the
$\forall$-player there exists a vertex $w_{i,j}$ pointing an edge of $G$.
So, the $\exists$-player has a win.

Assume now that $\exists$-player has a winning strategy. We prove
that the formula $Q$ is true. If the $\exists$-player traverses the
edges in (\ref{eq:seq1}) in order to reach $g_i$ then define
$x_i$ to be false, otherwise let the value of $x_i$ be set to true.
Then the $\forall$-player chooses any of the paths (\ref{eq:seq1})
or (\ref{eq:seq2}) which corresponds to setting an arbitrary Boolean
value to the variable $x_{i+1}$ quantified by $\forall$.
When the Boolean values have been assigned to the variables then,
by Fact~2, the $\forall$-player chooses a vertex $z_i$. Since the
$\exists$-player has a winning strategy, by Fact~3, at least one of the
vertices $w_{i,j}$, $j\in\{1,2,3\}$, is pointing an edge which still
belongs to $G$. By Fact~3, the literal $l_{i,j}$ of $F_i$ is true.
Since $i$ has been chosen arbitrarily, the formula $F$ is true.
\end{proof}

\section{Transformation of $G$ to the Phutball game}
\label{sec:graph_to_phutball}

\begin{figure}[htb]
\begin{center}
\begin{picture}(0,0)%
\includegraphics{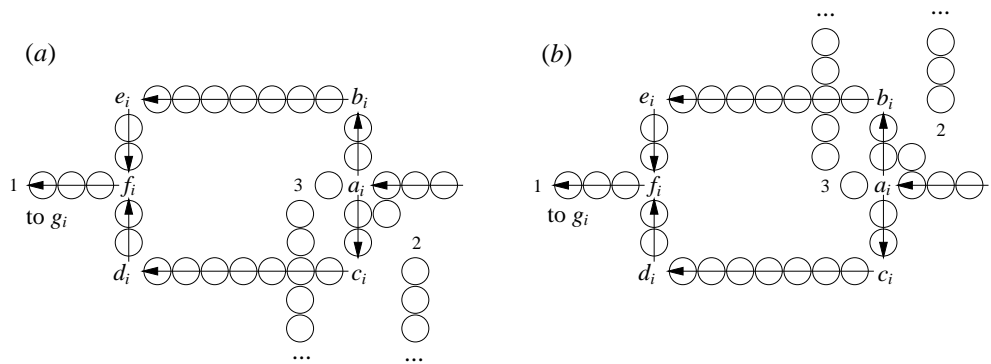}%
\end{picture}%
\setlength{\unitlength}{3158sp}%
\begingroup\makeatletter\ifx\SetFigFont\undefined%
\gdef\SetFigFont#1#2#3#4#5{%
  \reset@font\fontsize{#1}{#2pt}%
  \fontfamily{#3}\fontseries{#4}\fontshape{#5}%
  \selectfont}%
\fi\endgroup%
\begin{picture}(7717,2764)(101,-1828)
\put(3281,-932){\makebox(0,0)[lb]{\smash{{\SetFigFont{7}{8.4}{\familydefault}{\mddefault}{\updefault}{\color[rgb]{0,0,0}2}%
}}}}
\put(7349,906){\makebox(0,0)[lb]{\smash{{\SetFigFont{10}{12.0}{\familydefault}{\mddefault}{\updefault}{\color[rgb]{0,0,0}...}%
}}}}
\put(3218,-1813){\makebox(0,0)[lb]{\smash{{\SetFigFont{10}{12.0}{\familydefault}{\mddefault}{\updefault}{\color[rgb]{0,0,0}...}%
}}}}
\put(7396,-46){\makebox(0,0)[lb]{\smash{{\SetFigFont{7}{8.4}{\familydefault}{\mddefault}{\updefault}{\color[rgb]{0,0,0}2}%
}}}}
\put(2335,-1804){\makebox(0,0)[lb]{\smash{{\SetFigFont{10}{12.0}{\familydefault}{\mddefault}{\updefault}{\color[rgb]{0,0,0}...}%
}}}}
\put(2351,-480){\makebox(0,0)[lb]{\smash{{\SetFigFont{7}{8.4}{\familydefault}{\mddefault}{\updefault}{\color[rgb]{0,0,0}3}%
}}}}
\put(6449,897){\makebox(0,0)[lb]{\smash{{\SetFigFont{10}{12.0}{\familydefault}{\mddefault}{\updefault}{\color[rgb]{0,0,0}...}%
}}}}
\put(6488,-489){\makebox(0,0)[lb]{\smash{{\SetFigFont{7}{8.4}{\familydefault}{\mddefault}{\updefault}{\color[rgb]{0,0,0}3}%
}}}}
\put(5121,-491){\makebox(0,0)[lb]{\smash{{\SetFigFont{9}{10.8}{\familydefault}{\mddefault}{\updefault}{\color[rgb]{0,0,0}$f_i$}%
}}}}
\put(5057,-1182){\makebox(0,0)[lb]{\smash{{\SetFigFont{9}{10.8}{\familydefault}{\mddefault}{\updefault}{\color[rgb]{0,0,0}$d_i$}%
}}}}
\put(6936,-1186){\makebox(0,0)[lb]{\smash{{\SetFigFont{9}{10.8}{\familydefault}{\mddefault}{\updefault}{\color[rgb]{0,0,0}$c_i$}%
}}}}
\put(992,-487){\makebox(0,0)[lb]{\smash{{\SetFigFont{9}{10.8}{\familydefault}{\mddefault}{\updefault}{\color[rgb]{0,0,0}$f_i$}%
}}}}
\put(2802,-1181){\makebox(0,0)[lb]{\smash{{\SetFigFont{9}{10.8}{\familydefault}{\mddefault}{\updefault}{\color[rgb]{0,0,0}$c_i$}%
}}}}
\put(244,539){\makebox(0,0)[lb]{\smash{{\SetFigFont{10}{12.0}{\familydefault}{\mddefault}{\updefault}{\color[rgb]{0,0,0}($a$)}%
}}}}
\put(4294,530){\makebox(0,0)[lb]{\smash{{\SetFigFont{10}{12.0}{\familydefault}{\mddefault}{\updefault}{\color[rgb]{0,0,0}($b$)}%
}}}}
\put(6927,191){\makebox(0,0)[lb]{\smash{{\SetFigFont{9}{10.8}{\familydefault}{\mddefault}{\updefault}{\color[rgb]{0,0,0}$b_i$}%
}}}}
\put(2802,201){\makebox(0,0)[lb]{\smash{{\SetFigFont{9}{10.8}{\familydefault}{\mddefault}{\updefault}{\color[rgb]{0,0,0}$b_i$}%
}}}}
\put(946,206){\makebox(0,0)[lb]{\smash{{\SetFigFont{9}{10.8}{\familydefault}{\mddefault}{\updefault}{\color[rgb]{0,0,0}$e_i$}%
}}}}
\put(5062,213){\makebox(0,0)[lb]{\smash{{\SetFigFont{9}{10.8}{\familydefault}{\mddefault}{\updefault}{\color[rgb]{0,0,0}$e_i$}%
}}}}
\put(936,-1183){\makebox(0,0)[lb]{\smash{{\SetFigFont{9}{10.8}{\familydefault}{\mddefault}{\updefault}{\color[rgb]{0,0,0}$d_i$}%
}}}}
\put(251,-737){\makebox(0,0)[lb]{\smash{{\SetFigFont{9}{10.8}{\familydefault}{\mddefault}{\updefault}{\color[rgb]{0,0,0}to $g_i$}%
}}}}
\put(4343,-724){\makebox(0,0)[lb]{\smash{{\SetFigFont{9}{10.8}{\familydefault}{\mddefault}{\updefault}{\color[rgb]{0,0,0}to $g_i$}%
}}}}
\put(6910,-486){\makebox(0,0)[lb]{\smash{{\SetFigFont{9}{10.8}{\familydefault}{\mddefault}{\updefault}{\color[rgb]{0,0,0}$a_i$}%
}}}}
\put(2779,-491){\makebox(0,0)[lb]{\smash{{\SetFigFont{9}{10.8}{\familydefault}{\mddefault}{\updefault}{\color[rgb]{0,0,0}$a_i$}%
}}}}
\put(116,-477){\makebox(0,0)[lb]{\smash{{\SetFigFont{7}{8.4}{\familydefault}{\mddefault}{\updefault}{\color[rgb]{0,0,0}1}%
}}}}
\put(4231,-487){\makebox(0,0)[lb]{\smash{{\SetFigFont{7}{8.4}{\familydefault}{\mddefault}{\updefault}{\color[rgb]{0,0,0}1}%
}}}}
\end{picture}%
\caption{The variable component $G_F(x_i)$ for
         ($a$) $i=1,3,5,\ldots,n-1$, and
         ($b$) $i=2,4,6,\ldots,n-2$}
\label{pic:var_comp_transform}
\end{center}
\end{figure}

In the following we transform the input to the graph game, i.e.
a directed graph $G$, a set $C\subseteq V$, a starting vertex $s$
and a relation $R$, defined in
Section~\ref{sec:graph_game}, into a configuration of stones of the
Phutball game. Note that we do not give a reduction between
the two problems, but we only show how to code a
well structured instances of the graph game. This, together with
Theorem~\ref{thm:graph_game_pspace}, will give a desired reduction
from the QBF problem to the Phutball game. For brevity we will use
the symbols from the previous section used to denote the vertices of $G$
to refer to the points on the board (see e.g. Fig.~\ref{pic:var_comp_transform}($a$)).
Only the vertices in $C$ will be coded using special gadgets.
Because of the direct correspondence between the vertices of $G$
and the fields on the Phutball board we will use the labels
used for the vertices to denote the fields. It will be clear from
the context whether we refer to a vertex or to a point on the board.

\begin{figure}[htb]
\begin{center}
\begin{picture}(0,0)%
\includegraphics{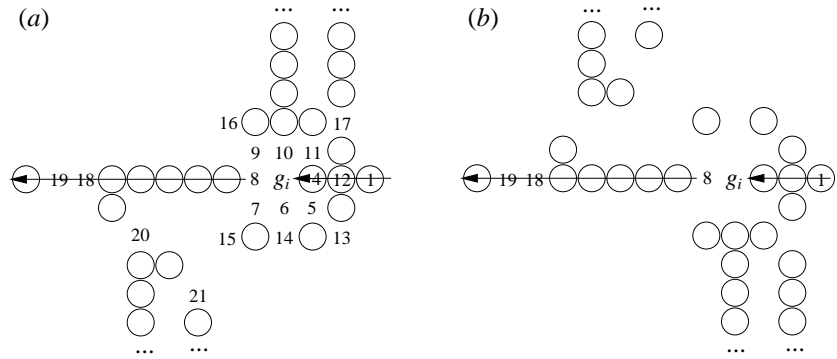}%
\end{picture}%
\setlength{\unitlength}{3158sp}%
\begingroup\makeatletter\ifx\SetFigFont\undefined%
\gdef\SetFigFont#1#2#3#4#5{%
  \reset@font\fontsize{#1}{#2pt}%
  \fontfamily{#3}\fontseries{#4}\fontshape{#5}%
  \selectfont}%
\fi\endgroup%
\begin{picture}(6517,2747)(-702,-2049)
\put(1402,653){\makebox(0,0)[lb]{\smash{{\SetFigFont{10}{12.0}{\familydefault}{\mddefault}{\updefault}{\color[rgb]{0,0,0}...}%
}}}}
\put(1217,-718){\makebox(0,0)[lb]{\smash{{\SetFigFont{7}{8.4}{\familydefault}{\mddefault}{\updefault}{\color[rgb]{0,0,0}8}%
}}}}
\put(1668,-943){\makebox(0,0)[lb]{\smash{{\SetFigFont{7}{8.4}{\familydefault}{\mddefault}{\updefault}{\color[rgb]{0,0,0}5}%
}}}}
\put(1449,-943){\makebox(0,0)[lb]{\smash{{\SetFigFont{7}{8.4}{\familydefault}{\mddefault}{\updefault}{\color[rgb]{0,0,0}6}%
}}}}
\put(1224,-950){\makebox(0,0)[lb]{\smash{{\SetFigFont{7}{8.4}{\familydefault}{\mddefault}{\updefault}{\color[rgb]{0,0,0}7}%
}}}}
\put(312,-2034){\makebox(0,0)[lb]{\smash{{\SetFigFont{10}{12.0}{\familydefault}{\mddefault}{\updefault}{\color[rgb]{0,0,0}...}%
}}}}
\put(740,-2021){\makebox(0,0)[lb]{\smash{{\SetFigFont{10}{12.0}{\familydefault}{\mddefault}{\updefault}{\color[rgb]{0,0,0}...}%
}}}}
\put(4955,-2026){\makebox(0,0)[lb]{\smash{{\SetFigFont{10}{12.0}{\familydefault}{\mddefault}{\updefault}{\color[rgb]{0,0,0}...}%
}}}}
\put(3825,647){\makebox(0,0)[lb]{\smash{{\SetFigFont{10}{12.0}{\familydefault}{\mddefault}{\updefault}{\color[rgb]{0,0,0}...}%
}}}}
\put(4292,661){\makebox(0,0)[lb]{\smash{{\SetFigFont{10}{12.0}{\familydefault}{\mddefault}{\updefault}{\color[rgb]{0,0,0}...}%
}}}}
\put(5417,-2016){\makebox(0,0)[lb]{\smash{{\SetFigFont{10}{12.0}{\familydefault}{\mddefault}{\updefault}{\color[rgb]{0,0,0}...}%
}}}}
\put(1409,-506){\makebox(0,0)[lb]{\smash{{\SetFigFont{7}{8.4}{\familydefault}{\mddefault}{\updefault}{\color[rgb]{0,0,0}10}%
}}}}
\put(1637,-507){\makebox(0,0)[lb]{\smash{{\SetFigFont{7}{8.4}{\familydefault}{\mddefault}{\updefault}{\color[rgb]{0,0,0}11}%
}}}}
\put(1866,-1182){\makebox(0,0)[lb]{\smash{{\SetFigFont{7}{8.4}{\familydefault}{\mddefault}{\updefault}{\color[rgb]{0,0,0}13}%
}}}}
\put(1413,-1179){\makebox(0,0)[lb]{\smash{{\SetFigFont{7}{8.4}{\familydefault}{\mddefault}{\updefault}{\color[rgb]{0,0,0}14}%
}}}}
\put(964,-1189){\makebox(0,0)[lb]{\smash{{\SetFigFont{7}{8.4}{\familydefault}{\mddefault}{\updefault}{\color[rgb]{0,0,0}15}%
}}}}
\put(976,-276){\makebox(0,0)[lb]{\smash{{\SetFigFont{7}{8.4}{\familydefault}{\mddefault}{\updefault}{\color[rgb]{0,0,0}16}%
}}}}
\put(1867,-280){\makebox(0,0)[lb]{\smash{{\SetFigFont{7}{8.4}{\familydefault}{\mddefault}{\updefault}{\color[rgb]{0,0,0}17}%
}}}}
\put(285,-1152){\makebox(0,0)[lb]{\smash{{\SetFigFont{7}{8.4}{\familydefault}{\mddefault}{\updefault}{\color[rgb]{0,0,0}20}%
}}}}
\put(740,-1636){\makebox(0,0)[lb]{\smash{{\SetFigFont{7}{8.4}{\familydefault}{\mddefault}{\updefault}{\color[rgb]{0,0,0}21}%
}}}}
\put(3166,-721){\makebox(0,0)[lb]{\smash{{\SetFigFont{7}{8.4}{\familydefault}{\mddefault}{\updefault}{\color[rgb]{0,0,0}19}%
}}}}
\put(3376,-721){\makebox(0,0)[lb]{\smash{{\SetFigFont{7}{8.4}{\familydefault}{\mddefault}{\updefault}{\color[rgb]{0,0,0}18}%
}}}}
\put(-356,-722){\makebox(0,0)[lb]{\smash{{\SetFigFont{7}{8.4}{\familydefault}{\mddefault}{\updefault}{\color[rgb]{0,0,0}19}%
}}}}
\put(4771,-710){\makebox(0,0)[lb]{\smash{{\SetFigFont{7}{8.4}{\familydefault}{\mddefault}{\updefault}{\color[rgb]{0,0,0}8}%
}}}}
\put(-148,-723){\makebox(0,0)[lb]{\smash{{\SetFigFont{7}{8.4}{\familydefault}{\mddefault}{\updefault}{\color[rgb]{0,0,0}18}%
}}}}
\put(-599,539){\makebox(0,0)[lb]{\smash{{\SetFigFont{10}{12.0}{\familydefault}{\mddefault}{\updefault}{\color[rgb]{0,0,0}($a$)}%
}}}}
\put(2926,539){\makebox(0,0)[lb]{\smash{{\SetFigFont{10}{12.0}{\familydefault}{\mddefault}{\updefault}{\color[rgb]{0,0,0}($b$)}%
}}}}
\put(1856,653){\makebox(0,0)[lb]{\smash{{\SetFigFont{10}{12.0}{\familydefault}{\mddefault}{\updefault}{\color[rgb]{0,0,0}...}%
}}}}
\put(1399,-716){\makebox(0,0)[lb]{\smash{{\SetFigFont{9}{10.8}{\familydefault}{\mddefault}{\updefault}{\color[rgb]{0,0,0}$g_i$}%
}}}}
\put(4953,-716){\makebox(0,0)[lb]{\smash{{\SetFigFont{9}{10.8}{\familydefault}{\mddefault}{\updefault}{\color[rgb]{0,0,0}$g_i$}%
}}}}
\put(2131,-720){\makebox(0,0)[lb]{\smash{{\SetFigFont{7}{8.4}{\familydefault}{\mddefault}{\updefault}{\color[rgb]{0,0,0}1}%
}}}}
\put(1866,-720){\makebox(0,0)[lb]{\smash{{\SetFigFont{7}{8.4}{\familydefault}{\mddefault}{\updefault}{\color[rgb]{0,0,0}12}%
}}}}
\put(5667,-720){\makebox(0,0)[lb]{\smash{{\SetFigFont{7}{8.4}{\familydefault}{\mddefault}{\updefault}{\color[rgb]{0,0,0}1}%
}}}}
\put(1706,-715){\makebox(0,0)[lb]{\smash{{\SetFigFont{7}{8.4}{\familydefault}{\mddefault}{\updefault}{\color[rgb]{0,0,0}4}%
}}}}
\put(1225,-513){\makebox(0,0)[lb]{\smash{{\SetFigFont{7}{8.4}{\familydefault}{\mddefault}{\updefault}{\color[rgb]{0,0,0}9}%
}}}}
\end{picture}%
\caption{Configuration of white stones corresponding to vertices
         ($a$) $g_i$, where $i$ is odd,
         ($b$) $g_i$, where $i$ is even}
\label{pic:switch}
\end{center}
\end{figure}

Let the upper (respectively lower) edge of the board be the
$\forall$-player's ($\exists$-player's, resp.) goal line.
The vertices in $V(G)\setminus C$ are coded as the empty points
on the board. We will choose those empty points in such a way that
if there is an edge $(u,v)\in E(G)$ then the points corresponding
to $u$ and $v$ will have the same horizontal or vertical coordinates.
The edges of the graph correspond to the (horizontal or vertical) sequences
of stones. The starting vertex is also coded as an empty spot and
it initially contains the ball.
The configuration of stones corresponding to the variable component $G(x_i)$ for
$i=1,3,5,\ldots,n-1$ is given in Fig.~\ref{pic:var_comp_transform}($a$)
while the Fig.~\ref{pic:var_comp_transform}($b$) gives the variable
component for $i=2,4,6,\ldots,n$.
In all the figures of this section, the dots ending a vertical line of
stones indicate that the line ends at the appropriate (upper or lower) goal line.
We will use two types of configurations corresponding to the
vertices $g_i$, $0<i<n$. Fig.~\ref{pic:switch}($a$) (Fig.~\ref{pic:switch}($b$))
presents the configuration corresponding to $g_i$ for
$i=1,3,5,\ldots,n-1$ ($i=2,4,6,\ldots,n-2$, respectively).
In order to make the analysis consistent
we will use the label $g_i$, $0<i<n$, to mark a point on the board
as shown in Fig.~\ref{pic:switch}.
Roughly speaking, such a gadget forces the following sequence of events:
one player makes a jump ending at field $g_i$ (according to an arrow on the
right hand side), then two stones are placed
at points $8$ and $18$ (each by one player) and finally the other player
makes a move of two jumps (as indicated by the second arrow) leading
the ball directly to the point $a_{i+1}$.
To obtain the configuration of stones corresponding to $G(x_i)$,
denoted in the following by $G_P(x_i)$,
for $i=1,3,5,\ldots,n-1$ (respectively for $i=2,4,6,\ldots,n-2$)
we connect the gadgets in Figs~\ref{pic:var_comp_transform}($a$)
and~\ref{pic:switch}($a$) (Figs~\ref{pic:var_comp_transform}($b$)
and~\ref{pic:switch}($b$), resp.) in such a way that the points
marked by $1$ in both pictures refer to the same place on the board.

We will use the following correspondence between the edges of $G$
and the lines of stones in $G_P$: an edge of $G$
\[(x,y)\in\{(g_0,a_1),(g_n,x_1)\}\cup\bigcup_{1\leq i\leq n} E(G(x_i))\setminus\{(g_{i},a_{i+1})\}\]
corresponds to a line of white stones between the points $x$ and $y$
on the board, while $(g_{i},a_{i+1})$, $i=1,\ldots,n-1$,
corresponds to two lines of stones: between $g_i$ and $19$ of $G_P(x_{i})$
and between $19$ of $G_P(x_{i})$ and $a_{i+1}$ of $G_P(x_{i+1})$.
Note that two points in a line between $g_i$ and $19$ are by the
definition unoccupied, but the game is set up in such a way that
when the ball is about to move from $g_i$ then there is a line
of white stones between $g_i$ and $19$ of $G_P(x_i)$ which we are
going to prove later.

Note that we used the Facts 1-3 listed in the previous section
to obtain Theorem~\ref{thm:graph_game_pspace}.
Now we prove these facts for the corresponding configurations of stones
on the board. Then we may conclude that the game of Phutball simulates
the graph game which will give us a desired reduction from the QBF
problem. If $x$ and $y$ are two points on the board then
$x\to y$ denotes a jump from $x$ to $y$ and removing all the
stones between $x$ and $y$ (we will use this symbol in such a way
that all the conditions required by the rules of the game for making
a jump will be satisfied).

\textbf{Proof of Fact 1:}
In the terms of the phutball game we are going to prove, by an induction on $i$,
that if a ball is at $g_{i-1}$, $1<i\leq n$, and is $X$ the player making the
next move then the following sequence of moves occurs:
\begin{list}{}{}
 \item[(i)] $X$ places a white stone at $8$ of $G_P(x_{i-1})$,
 \item[(ii)] $\overline{X}$ places a white stone at $18$ of $G_P(x_{i-1})$,
 \item[(iii)] $X$ makes a sequence of jumps over the lines of stones corresponding to the edges given in~(\ref{eq:seq1}) or~(\ref{eq:seq2}).
\end{list}
Moveover, $X$ is the $\exists$-player ($\forall$-player) for
odd (even, respectively) values of $i$.
For $i=1$ the situation is similar to the case when $i>1$ except that only (iii) is done.

So, assume that the ball is at $g_{i-1}$. By the induction hypothesis, the
white stones on the right hand side of $g_{i-1}$ are no longer on the board.
Moreover, both for odd and even values of $i$, the lines of stones next to
$10$ and $17$ ($20$ and $21$) of $G_F(x_{i-1})$ lead to the $X$'s
($\overline{X}$'s, respectively) board line.
We have that $X$ must place a white stone, because he cannot make a jump.
Clearly, he cannot put a stone at $10$ of $G_P(x_{i-1})$. 
If he does not occupy one of the fields $4$-$11$ of $G_P(x_{i-1})$
then in the next turn $\overline{X}$ puts a white stone at the point $10$ of $G_P(x_{i-1})$
and it is easy to see that $\overline{X}$ wins the game. Observe, that if $\overline{X}$
is able to move the ball to the point $17$ then he is one jump away from
his opponent's goal line. Note that if $X$ places a stone in one of the points
$4,5,6,7,9,11$ of $G_P(x_{i-1})$ then $\overline{X}$ can reach $17$
and win the game. Thus, $X$ places a white stone at the field $8$ of $G_P(x_{i-1})$,
i.e. (i) occurs.
Then, $\overline{X}$ can either:
(1) make one of the moves $g_{i-1}\to 18$, $g_{i-1}\to 18\to 20$ or $g_{i-1}\to 18\to 20\to 21$,
    but it is easy to see that in all cases his opponent wins in the next turn, or
(2) put a white stone and if he chooses a field different than $18$ then,
    similarly as in (1), he loses, because either $20$ or $21$ of $G_F(x_{i-1})$ is
    unoccupied. So, $X$ can make jumps $g_{i-1}\to 18\to 20$
    or $g_{i-1}\to 18\to 21$, respectively and reach his opponent's goal line in the next jump.
This proves that (ii) must happen.

The fact that $X$ must reach the point $g_i$ by jumping over the lines
of stones corresponding to~(\ref{eq:seq1}) or~(\ref{eq:seq2}) follows
from the observation that otherwise he loses the game.
In particular, if $X$ places a white stone somewhere on the board
instead of making some jumps then one of the points $3$ or $a_i$ of
$G_P(x_{i})$ is still unoccupied. Then, $\overline{X}$ reaches
$3$ or $2$ in $G_P(x_{i})$, respectively, and his next jump places the
ball at the $X$'s goal line.
So, $X$ is forced to make a sequence of jumps.
If $X$ finishes his move before reaching $b_i$ or $c_i$ then his moves were
$g_{i-1}\to 19$, $g_{i-1}\to 19\to a_i$, $g_{i-1}\to 19\to a_i\to 2$
or $g_{i-1}\to 19\to a_i\to 3$ and it is easy to see that $\overline{X}$
wins immediately in all cases.
If $X$ reaches $b_i$, by jumps $g_{i-1}\to 19\to a_i\to b_i$
(the case of $c_i$ is analogous) then he must follow to the point $g_i$
(by jumps $b_i\to e_i\to f_i\to g_i$) since otherwise there is a path
$b_i\to e_i\to d_i\to c_i\to a_i\to 2$ which $\overline{X}$ can follow.
$\Box$

\begin{figure}[htb]
\begin{center}
\begin{picture}(0,0)%
\includegraphics{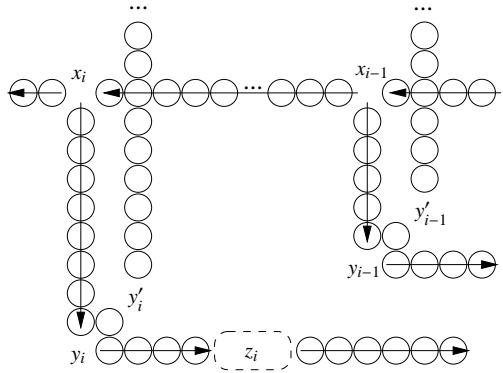}%
\end{picture}%
\setlength{\unitlength}{3158sp}%
\begingroup\makeatletter\ifx\SetFigFont\undefined%
\gdef\SetFigFont#1#2#3#4#5{%
  \reset@font\fontsize{#1}{#2pt}%
  \fontfamily{#3}\fontseries{#4}\fontshape{#5}%
  \selectfont}%
\fi\endgroup%
\begin{picture}(3923,2877)(-86,-1873)
\put(901,974){\makebox(0,0)[lb]{\smash{{\SetFigFont{10}{12.0}{\familydefault}{\mddefault}{\updefault}{\color[rgb]{0,0,0}...}%
}}}}
\put(3137,970){\makebox(0,0)[lb]{\smash{{\SetFigFont{10}{12.0}{\familydefault}{\mddefault}{\updefault}{\color[rgb]{0,0,0}...}%
}}}}
\put(1800,341){\makebox(0,0)[lb]{\smash{{\SetFigFont{10}{12.0}{\familydefault}{\mddefault}{\updefault}{\color[rgb]{0,0,0}...}%
}}}}
\put(2680,457){\makebox(0,0)[lb]{\smash{{\SetFigFont{8}{9.6}{\familydefault}{\mddefault}{\updefault}{\color[rgb]{0,0,0}$x_{i-1}$}%
}}}}
\put(452,423){\makebox(0,0)[lb]{\smash{{\SetFigFont{8}{9.6}{\familydefault}{\mddefault}{\updefault}{\color[rgb]{0,0,0}$x_i$}%
}}}}
\put(1813,-1762){\makebox(0,0)[lb]{\smash{{\SetFigFont{8}{9.6}{\familydefault}{\mddefault}{\updefault}{\color[rgb]{0,0,0}$z_i$}%
}}}}
\put(451,-1786){\makebox(0,0)[lb]{\smash{{\SetFigFont{8}{9.6}{\familydefault}{\mddefault}{\updefault}{\color[rgb]{0,0,0}$y_i$}%
}}}}
\put(2626,-1111){\makebox(0,0)[lb]{\smash{{\SetFigFont{8}{9.6}{\familydefault}{\mddefault}{\updefault}{\color[rgb]{0,0,0}$y_{i-1}$}%
}}}}
\put(3151,-661){\makebox(0,0)[lb]{\smash{{\SetFigFont{8}{9.6}{\familydefault}{\mddefault}{\updefault}{\color[rgb]{0,0,0}$y_{i-1}'$}%
}}}}
\put(901,-1336){\makebox(0,0)[lb]{\smash{{\SetFigFont{8}{9.6}{\familydefault}{\mddefault}{\updefault}{\color[rgb]{0,0,0}$y_i'$}%
}}}}
\end{picture}%
\caption{The configuration representing the paths in (\ref{eq:seq3})}
\label{pic:clause_transform}
\end{center}
\end{figure}

\begin{figure}[htb]
\begin{center}
\begin{picture}(0,0)%
\includegraphics{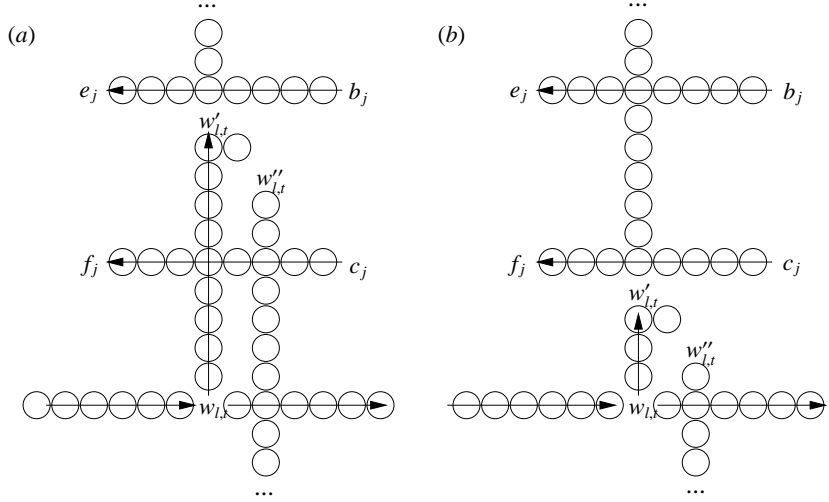}%
\end{picture}%
\setlength{\unitlength}{3158sp}%
\begingroup\makeatletter\ifx\SetFigFont\undefined%
\gdef\SetFigFont#1#2#3#4#5{%
  \reset@font\fontsize{#1}{#2pt}%
  \fontfamily{#3}\fontseries{#4}\fontshape{#5}%
  \selectfont}%
\fi\endgroup%
\begin{picture}(6477,3885)(2086,-1534)
\put(5476,2039){\makebox(0,0)[lb]{\smash{{\SetFigFont{9}{10.8}{\familydefault}{\mddefault}{\updefault}{\color[rgb]{0,0,0}($b$)}%
}}}}
\put(6968,2321){\makebox(0,0)[lb]{\smash{{\SetFigFont{10}{12.0}{\familydefault}{\mddefault}{\updefault}{\color[rgb]{0,0,0}...}%
}}}}
\put(7422,-1499){\makebox(0,0)[lb]{\smash{{\SetFigFont{10}{12.0}{\familydefault}{\mddefault}{\updefault}{\color[rgb]{0,0,0}...}%
}}}}
\put(2101,2039){\makebox(0,0)[lb]{\smash{{\SetFigFont{9}{10.8}{\familydefault}{\mddefault}{\updefault}{\color[rgb]{0,0,0}($a$)}%
}}}}
\put(3586,2315){\makebox(0,0)[lb]{\smash{{\SetFigFont{10}{12.0}{\familydefault}{\mddefault}{\updefault}{\color[rgb]{0,0,0}...}%
}}}}
\put(4032,-1519){\makebox(0,0)[lb]{\smash{{\SetFigFont{10}{12.0}{\familydefault}{\mddefault}{\updefault}{\color[rgb]{0,0,0}...}%
}}}}
\put(2664,1611){\makebox(0,0)[lb]{\smash{{\SetFigFont{9}{10.8}{\familydefault}{\mddefault}{\updefault}{\color[rgb]{0,0,0}$e_j$}%
}}}}
\put(2671,244){\makebox(0,0)[lb]{\smash{{\SetFigFont{9}{10.8}{\familydefault}{\mddefault}{\updefault}{\color[rgb]{0,0,0}$f_j$}%
}}}}
\put(6026,244){\makebox(0,0)[lb]{\smash{{\SetFigFont{9}{10.8}{\familydefault}{\mddefault}{\updefault}{\color[rgb]{0,0,0}$f_j$}%
}}}}
\put(6040,1610){\makebox(0,0)[lb]{\smash{{\SetFigFont{9}{10.8}{\familydefault}{\mddefault}{\updefault}{\color[rgb]{0,0,0}$e_j$}%
}}}}
\put(4781,224){\makebox(0,0)[lb]{\smash{{\SetFigFont{9}{10.8}{\familydefault}{\mddefault}{\updefault}{\color[rgb]{0,0,0}$c_j$}%
}}}}
\put(4775,1590){\makebox(0,0)[lb]{\smash{{\SetFigFont{9}{10.8}{\familydefault}{\mddefault}{\updefault}{\color[rgb]{0,0,0}$b_j$}%
}}}}
\put(8187,1598){\makebox(0,0)[lb]{\smash{{\SetFigFont{9}{10.8}{\familydefault}{\mddefault}{\updefault}{\color[rgb]{0,0,0}$b_j$}%
}}}}
\put(8182,253){\makebox(0,0)[lb]{\smash{{\SetFigFont{9}{10.8}{\familydefault}{\mddefault}{\updefault}{\color[rgb]{0,0,0}$c_j$}%
}}}}
\put(3601,-886){\makebox(0,0)[lb]{\smash{{\SetFigFont{9}{10.8}{\familydefault}{\mddefault}{\updefault}{\color[rgb]{0,0,0}$w_{l,t}$}%
}}}}
\put(6976,-886){\makebox(0,0)[lb]{\smash{{\SetFigFont{9}{10.8}{\familydefault}{\mddefault}{\updefault}{\color[rgb]{0,0,0}$w_{l,t}$}%
}}}}
\put(3601,1364){\makebox(0,0)[lb]{\smash{{\SetFigFont{9}{10.8}{\familydefault}{\mddefault}{\updefault}{\color[rgb]{0,0,0}$w_{l,t}'$}%
}}}}
\put(4051,914){\makebox(0,0)[lb]{\smash{{\SetFigFont{9}{10.8}{\familydefault}{\mddefault}{\updefault}{\color[rgb]{0,0,0}$w_{l,t}''$}%
}}}}
\put(6976, 14){\makebox(0,0)[lb]{\smash{{\SetFigFont{9}{10.8}{\familydefault}{\mddefault}{\updefault}{\color[rgb]{0,0,0}$w_{l,t}'$}%
}}}}
\put(7426,-436){\makebox(0,0)[lb]{\smash{{\SetFigFont{9}{10.8}{\familydefault}{\mddefault}{\updefault}{\color[rgb]{0,0,0}$w_{l,t}''$}%
}}}}
\end{picture}%
\caption{($a$) $w_{l,2}$ and pointing to $(b_j,e_j)$
         and ($b$) $w_{l,3}$ and pointing to $(e_i,f_i)$}
\label{pic:pointing_edge}
\end{center}
\end{figure}

\begin{figure}[htb]
\begin{center}
\begin{picture}(0,0)%
\includegraphics{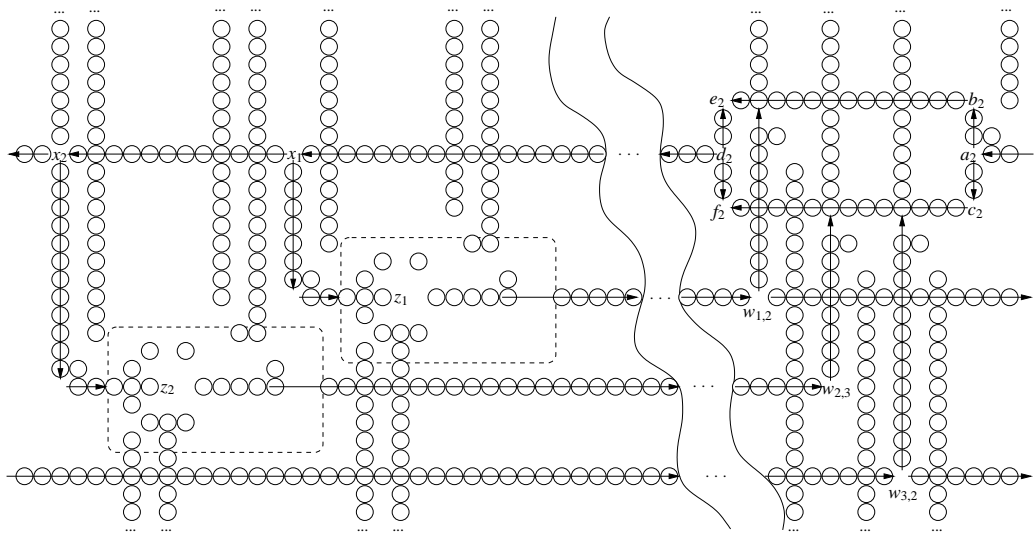}%
\end{picture}%
\setlength{\unitlength}{1973sp}%
\begingroup\makeatletter\ifx\SetFigFont\undefined%
\gdef\SetFigFont#1#2#3#4#5{%
  \reset@font\fontsize{#1}{#2pt}%
  \fontfamily{#3}\fontseries{#4}\fontshape{#5}%
  \selectfont}%
\fi\endgroup%
\begin{picture}(12924,6583)(-7361,-3764)
\put(-4285,2789){\makebox(0,0)[lb]{\smash{{\SetFigFont{6}{7.2}{\familydefault}{\mddefault}{\updefault}{\color[rgb]{0,0,0}...}%
}}}}
\put(-4735,2789){\makebox(0,0)[lb]{\smash{{\SetFigFont{6}{7.2}{\familydefault}{\mddefault}{\updefault}{\color[rgb]{0,0,0}...}%
}}}}
\put(-5410,-3749){\makebox(0,0)[lb]{\smash{{\SetFigFont{6}{7.2}{\familydefault}{\mddefault}{\updefault}{\color[rgb]{0,0,0}...}%
}}}}
\put(-5860,-3749){\makebox(0,0)[lb]{\smash{{\SetFigFont{6}{7.2}{\familydefault}{\mddefault}{\updefault}{\color[rgb]{0,0,0}...}%
}}}}
\put(-2497,-3743){\makebox(0,0)[lb]{\smash{{\SetFigFont{6}{7.2}{\familydefault}{\mddefault}{\updefault}{\color[rgb]{0,0,0}...}%
}}}}
\put(-2947,-3743){\makebox(0,0)[lb]{\smash{{\SetFigFont{6}{7.2}{\familydefault}{\mddefault}{\updefault}{\color[rgb]{0,0,0}...}%
}}}}
\put(-1810,2789){\makebox(0,0)[lb]{\smash{{\SetFigFont{6}{7.2}{\familydefault}{\mddefault}{\updefault}{\color[rgb]{0,0,0}...}%
}}}}
\put(-1360,2789){\makebox(0,0)[lb]{\smash{{\SetFigFont{6}{7.2}{\familydefault}{\mddefault}{\updefault}{\color[rgb]{0,0,0}...}%
}}}}
\put(5143,2784){\makebox(0,0)[lb]{\smash{{\SetFigFont{6}{7.2}{\familydefault}{\mddefault}{\updefault}{\color[rgb]{0,0,0}...}%
}}}}
\put(3810,2777){\makebox(0,0)[lb]{\smash{{\SetFigFont{6}{7.2}{\familydefault}{\mddefault}{\updefault}{\color[rgb]{0,0,0}...}%
}}}}
\put(2911,2770){\makebox(0,0)[lb]{\smash{{\SetFigFont{6}{7.2}{\familydefault}{\mddefault}{\updefault}{\color[rgb]{0,0,0}...}%
}}}}
\put(1998,2770){\makebox(0,0)[lb]{\smash{{\SetFigFont{6}{7.2}{\familydefault}{\mddefault}{\updefault}{\color[rgb]{0,0,0}...}%
}}}}
\put(2471,-3749){\makebox(0,0)[lb]{\smash{{\SetFigFont{6}{7.2}{\familydefault}{\mddefault}{\updefault}{\color[rgb]{0,0,0}...}%
}}}}
\put(3371,-3743){\makebox(0,0)[lb]{\smash{{\SetFigFont{6}{7.2}{\familydefault}{\mddefault}{\updefault}{\color[rgb]{0,0,0}...}%
}}}}
\put(4271,-3743){\makebox(0,0)[lb]{\smash{{\SetFigFont{6}{7.2}{\familydefault}{\mddefault}{\updefault}{\color[rgb]{0,0,0}...}%
}}}}
\put(-6757,2769){\makebox(0,0)[lb]{\smash{{\SetFigFont{6}{7.2}{\familydefault}{\mddefault}{\updefault}{\color[rgb]{0,0,0}...}%
}}}}
\put(-6310,2776){\makebox(0,0)[lb]{\smash{{\SetFigFont{6}{7.2}{\familydefault}{\mddefault}{\updefault}{\color[rgb]{0,0,0}...}%
}}}}
\put(-3371,2783){\makebox(0,0)[lb]{\smash{{\SetFigFont{6}{7.2}{\familydefault}{\mddefault}{\updefault}{\color[rgb]{0,0,0}...}%
}}}}
\put(1433,-3070){\makebox(0,0)[lb]{\smash{{\SetFigFont{6}{7.2}{\familydefault}{\mddefault}{\updefault}{\color[rgb]{0,0,0}. . .}%
}}}}
\put(1274,-1943){\makebox(0,0)[lb]{\smash{{\SetFigFont{6}{7.2}{\familydefault}{\mddefault}{\updefault}{\color[rgb]{0,0,0}. . .}%
}}}}
\put(733,-830){\makebox(0,0)[lb]{\smash{{\SetFigFont{6}{7.2}{\familydefault}{\mddefault}{\updefault}{\color[rgb]{0,0,0}. . .}%
}}}}
\put(340,970){\makebox(0,0)[lb]{\smash{{\SetFigFont{6}{7.2}{\familydefault}{\mddefault}{\updefault}{\color[rgb]{0,0,0}. . .}%
}}}}
\put(3736,-3331){\makebox(0,0)[lb]{\smash{{\SetFigFont{7}{8.4}{\familydefault}{\mddefault}{\updefault}{\color[rgb]{0,0,0}$w_{3,2}$}%
}}}}
\put(1576,914){\makebox(0,0)[lb]{\smash{{\SetFigFont{7}{8.4}{\familydefault}{\mddefault}{\updefault}{\color[rgb]{0,0,0}$d_2$}%
}}}}
\put(4741,1604){\makebox(0,0)[lb]{\smash{{\SetFigFont{7}{8.4}{\familydefault}{\mddefault}{\updefault}{\color[rgb]{0,0,0}$b_2$}%
}}}}
\put(-6784,919){\makebox(0,0)[lb]{\smash{{\SetFigFont{7}{8.4}{\familydefault}{\mddefault}{\updefault}{\color[rgb]{0,0,0}$x_2$}%
}}}}
\put(-3824,924){\makebox(0,0)[lb]{\smash{{\SetFigFont{7}{8.4}{\familydefault}{\mddefault}{\updefault}{\color[rgb]{0,0,0}$x_1$}%
}}}}
\put(-5409,-1996){\makebox(0,0)[lb]{\smash{{\SetFigFont{7}{8.4}{\familydefault}{\mddefault}{\updefault}{\color[rgb]{0,0,0}$z_2$}%
}}}}
\put(-2474,-876){\makebox(0,0)[lb]{\smash{{\SetFigFont{7}{8.4}{\familydefault}{\mddefault}{\updefault}{\color[rgb]{0,0,0}$z_1$}%
}}}}
\put(1486,1619){\makebox(0,0)[lb]{\smash{{\SetFigFont{7}{8.4}{\familydefault}{\mddefault}{\updefault}{\color[rgb]{0,0,0}$e_2$}%
}}}}
\put(1496,204){\makebox(0,0)[lb]{\smash{{\SetFigFont{7}{8.4}{\familydefault}{\mddefault}{\updefault}{\color[rgb]{0,0,0}$f_2$}%
}}}}
\put(1906,-1061){\makebox(0,0)[lb]{\smash{{\SetFigFont{7}{8.4}{\familydefault}{\mddefault}{\updefault}{\color[rgb]{0,0,0}$w_{1,2}$}%
}}}}
\put(2901,-2026){\makebox(0,0)[lb]{\smash{{\SetFigFont{7}{8.4}{\familydefault}{\mddefault}{\updefault}{\color[rgb]{0,0,0}$w_{2,3}$}%
}}}}
\put(4731,224){\makebox(0,0)[lb]{\smash{{\SetFigFont{7}{8.4}{\familydefault}{\mddefault}{\updefault}{\color[rgb]{0,0,0}$c_2$}%
}}}}
\put(4641,919){\makebox(0,0)[lb]{\smash{{\SetFigFont{7}{8.4}{\familydefault}{\mddefault}{\updefault}{\color[rgb]{0,0,0}$a_2$}%
}}}}
\end{picture}%
\caption{Some parts of the graph in Fig.~\ref{pic:graph_ex}}
\label{pic:board_ex}
\end{center}
\end{figure}

The conversion of the formula component $G(F)$ to the configuration of
stones is shown in Figs~\ref{pic:clause_transform} and \ref{pic:pointing_edge}.
In particular, Fig.~\ref{pic:clause_transform} depicts
the board representation of the edges $(x_{i-1},x_i)$,
$(x_i,y_i)$ and $(y_i,z_i)$ while Fig.~\ref{pic:pointing_edge}($a$)
(respectively Fig.~\ref{pic:pointing_edge}($b$)) gives
the configuration of stones coding the situation when $w_{i,t}$
is pointing an edge $(b_j,e_j)$ ($(c_j,f_j)$, respectively),
where $i\in\{1,\ldots,m\}$, $t\in\{1,2,3\}$ and $j\in\{1,\ldots,n\}$.
See Fig.~\ref{pic:board_ex} for an example where $G_P(x_2)$ is given together
with $w_{1,2}$ pointing $(b_2,e_2)$ and $w_{2,3}, w_{3,2}$ both
pointing $(c_2,f_2)$.
The configuration of stones corresponding to a vertex $z_i$
is identical to the one in Fig.~\ref{pic:switch}($a$), but it is
rotated with the angle of 180 degrees, and we use the symbol $z_i$
to refer to the point denoted by $g_i$ in the case of $G_P(x_i)$'s.
The configuration of stones corresponding to $G(F)$ is denoted by $G_P(F)$.

It remains to mention that the number of white stones in a line does
not change the analysis of the game. In particular the distance between $b_j$ and $e_j$
can be arbitrary long and in our reduction it depends on the number
of vertices $w_{i,t}$ pointing $(b_j,e_j)$ or $(c_j,f_j)$. We require
that there a distance of at least one 'field' between each pair of
vertical lines appearing on the board. In this way if one stone, say from a field $z$,
in a line between $x$ and $y$ has been removed from the board during the game then a jump
$x\to y$, where $y$ may be on the goal line, is replaced by two jumps
$x\to z\to y$. Thanks to the distance between the vertical lines,
no other jump from $z$ is possible. If a player ends his move at $z$ then
such a situation does not differ from the case when the move would end at
$y$ and our analysis covers that.

\textbf{Proof of Fact 2:}
From the Fact~1 it follows that when the ball is at point $g_n$ then
the $\forall$-player is the current player. The player cannot place
a stone on the board in the forthcoming move, because then the
$\exists$-player selects an empty point $x_i$ (it exists, because
we assumed w.l.o.g. that there are at least two variables in
the formula) and
\begin{equation}
 g_n\to x_1\to\cdots\to x_i\to y_i\to y_i'
 \label{eq:cannot_stay}
\end{equation}
leads directly to the $\forall$-player goal line.

By a simple induction on $i$ one can prove that when the
$\forall$-player reaches the point $x_i$ then he must either
jump to $x_{i+1}$ or to $y_i$. So, there exists $i\in\{1,\ldots,m\}$
such that the $\forall$-player reaches $y_i$. He cannot end his sequence of jumps
at $y_i$ or $y_i'$,
because, as before, he would immediately lose.
By the arguments similar to the ones presented in the proof of Fact~1,
the moves (i) and (ii) in the component corresponding to the vertex $z_i$ of $G$
must occur and the $\exists$-player begins his move when the ball is at
$z_i$.
$\Box$

\textbf{Proof of Fact 3:}
By Fact~2, the $\forall$-player reaches $z_i$. Similarly as in the proof of Fact~1
one can show that (i) and (ii), defined in the proof, occur and consequently
$\exists$-player starts his move from $z_i$ when the fields $8$ and $18$
are already occupied by white stones. The $\exists$-player is forced
to make a jump, because otherwise the $\forall$-player follows
a path 
\[z_i\to w_{i,1}\to\cdots\to w_{i,j}\to w_{i,j}'\to w_{i,j}''\]
(see
Fig.~\ref{pic:pointing_edge}), where $j\in\{1,2,3\}$ can be
chosen in such a way that none of $w_{i,j},w_{i,j}',w_{i,j}''$ is occupied
by a white stone placed by the $\exists$-player in his last move.
So, the $\exists$-player has three paths to follow:
\begin{equation}
z_i\to w_{i,1}\to\cdots\to w_{i,j}\to w_{i,j}',
\label{eq:three_paths}
\end{equation}
where $j\in\{1,2,3\}$. If he ends his move at $w_{i,j}'$ (or earlier) then
the $\forall$-player reaches $w_{i,j}''$ and can jump to the
$\exists$-player's board line. There are two cases to consider:
(i) $w_{i,j}$ points to an edge $(b_l,e_l)$;
(ii) $w_{i,j}$ points to an edge $(c_l,f_l)$.
Both cases are analogous, so we consider only (i). From the construction
of the board it follows that the only continuation of moves in
(\ref{eq:three_paths}) by the $\exists$-player in direction different than to $w_{i,j}''$ is possible when
no jump from $b_l$ to $e_l$ has been made during the game, and in this
case the $\exists$-player makes a jump from $w_{i,j}'$ to the
$\forall$-player's board line. So the $\exists$-player wins if and
only if no jump from $b_l$ to $e_l$ has been made (or equivalently the
edge $(b_l,e_l)$ still belongs to the corresponding graph $G$).
It remains to mention that the line of stones between $c_l$ and $f_l$
has been removed from the board, but the result is that a player makes
two jumps instead of one in order to move the ball from $w_{i,j}$ to $w_{i,j}'$
or from $w_{i,j}''$ to the board line.
$\Box$

Fig.~\ref{pic:board_ex} presents some parts of the board obtained
on the basis of the graph $G$ corresponding to formula
(\ref{eq:QBF_example}) and shown in Fig.~\ref{pic:graph_ex}.

\begin{theorem}
The game of Phutball is \textup{PSPACE}-hard.
\end{theorem}
\begin{proof}
The theorem follows from Facts~1-3, the proof of Theorem~\ref{thm:graph_game_pspace}
and an observation that the size of the board is polynomial in $n+m$.
\end{proof}

\section{Summary}

There are several natural questions one may ask about the complexity
of a game. One of them is: given an arbitrary state of the game, is it
possible for the current player to win in the next move? Such a
problem has been considered in \cite{phutball_endgames} where it has
been shown that it is NP-complete for Phutball. According to the discussion in
\cite{phutball_endgames} the games of Checkers and Phutball have many
similarities. However, it turns out that we can give a positive answer
to the above question in the case of Checkers in polynomial time
\cite{phutball_endgames,checkers_complexity}. Another question to ask
about the complexity of the game is the one considered in this
paper. In the case of Checkers, Fraenkel et. al. have shown that the
game is PSPACE-hard. Their result has been strenghtened by the paper
of Robson: the game is EXPTIME-complete \cite{checkers_exptime}. In
this paper we developed the first result concerning the complexity of
Phutball and an open question remains whether it belongs to PSPACE or
is as hard as EXPTIME?

\end{document}